\documentclass[singlecol]{epl2} 
\usepackage{graphicx}
\usepackage{amsmath}
\usepackage{amsfonts}
\usepackage{amssymb}
\newcommand{\ket}[1]{|#1\rangle}
\newcommand{\bra}[1]{\langle #1|}

\title{Magnetoresistance in Disordered Graphene: 
The Role of Pseudospin and Dimensionality Effects Unraveled}
\shorttitle{ Magnetoresistance in Disordered Graphene } 

\author{F. Ortmann\inst{1} \and A. Cresti\inst{2} \and G. Montambaux\inst{3} \and S. Roche\inst{4,5} }
\shortauthor{F. Ortmann \etal}

\institute{                    
  \inst{1} CEA, INAC, SPRAM, GT, 17 rue des Martyrs, 38054 Grenoble Cedex 9, France \\
\inst{2} IMEP-LAHC, UMR 5130 (Grenoble INP/UJF/CNRS/Universit\'{e} de Savoie), Minatec, 3 Parvis Louis N\'{e}el, 38016 Grenoble, France\\
\inst{3} Laboratoire de Physique des Solides, UMR 8502 du CNRS,
Universit\'{e} Paris-Sud, 91405 Orsay, France\\
\inst{4} CIN2 (ICN-CSIC) and Universitat Autonoma de Barcelona, Catalan Institute of Nanotechnology, Campus de la UAB, 08193 Bellaterra (Barcelona), Spain \\
\inst{5} ICREA, Institucio Catalana de Recerca i Estudis Avan\c cats, 08010 Barcelona, Spain}


\abstract{
We report a theoretical low-field magnetotransport study unveiling the effect of pseudospin in realistic models of weakly disordered graphene-based materials. Using an efficient Kubo computational method,  and simulating the effect of charges trapped in the oxide, different magnetoconductance fingerprints are numerically obtained in system sizes as large as $0.3 \mu\text{m}^{2}$, containing tens of millions of carbon atoms. In two-dimensional graphene,  a strong valley mixing  is found to irreparably yield a positive magnetoconductance (weak localization), whereas crossovers from positive to a negative magnetoconductance
 (weak antilocalization) are obtained by reducing disorder strength down to the ballistic limit. In sharp contrast, graphene nanoribbons with lateral size as large as 10nm show no sign of weak antilocalization, even for very small disorder strength. Our results rationalize the emergence of a complex phase diagram of magnetoconductance fingerprints, shedding some new light on the microscopical origin of pseudospin effects.
}
 
\begin{document}

\maketitle

In 2004, the report on graphene discovery~\cite{Novoselov} has sparked a great scientific excitement because of both novel type of electronic excitations (so-called {\it massless Dirac Fermions})~\cite{RMP} and a promising future of graphene-based technologies~\cite{Lemme}. In two-dimensional graphene, the very peculiar nature of low-energy electronic states (encompassing a new pseudospin degree of freedom) yields a wealth of anomalous transport features such as Klein tunneling~\cite{Klein},  weak antilocalization~\cite{WAL1,WAL2}, unconventional quantum Hall effect~\cite{QHE1,QHE2}, or new ways (supercollimation) to guide charge flows~\cite{Park} .  

Besides, the nature of disorder, its effect on electronic properties, and the envisionned defect engineering for novel graphene electronics are currently subjects of great concern~\cite{Bubble,Paco}. However to date, the precise relationship between the underlying disorder features and the onset of graphene unique transport properties remains debated. The question also applies to graphene nanoribbons which exhibit different electronic properties induced by lateral size confinement effects and edge symmetries~\cite{Ribbonpure1,Son,Son0,NLWhite,Biel}. 

One intriguing question is the precise role of pseudospin effects in the modulations of the conductance fingerprints in disordered  graphene based materials (both two-dimensional and nanoribbons).  Indeed, in ordinary low-dimensional disordered metals, quantum interferences between multiple scattering paths produce a quantum correction to the semiclassical (Drude) conductivity which can be either reduced (weak localization)~\cite{RMPLee} or enhanced (weak antilocalization-WAL)~\cite{Bergmanso}, depending on the strength of spin-orbit coupling.  In graphene, it was suggested that, even in absence of spin-orbit coupling, WAL develops if the underlying long range disorder preserves the pseudospin symmetry of the honeycomb geometry~\cite{WAL1,WAL2}. A perturbation (diagrammatic) theory was proposed, but to determine the quantum correction of conductivity, several phenomenological parameters were introduced (intravalley and intervalley elastic scattering times), out of reach from analytical considerations~\cite{WAL2}. This phenomenological approach provides a fitting equation for comparison with experimental data, but its connection to the specific microscopic disorder at the origin of multiple scattering remains totally elusive. 
 
A complex phase diagram of magnetoresistance phenomena in graphene has been recently experimentally revealed, exhibiting multiple crossovers from weak antilocalization (negative magnetoconductance) to weak localization (positive magnetoconductance) when tuning the energy of charge carriers or sample temperature~\cite{WALEX2}. The understanding of the microscopical origin of such a complex phase diagram and its relation to a possible metal-insulator transition remains however obscure. Its comprehension demands to clarify the connection between the precise disorder characteristics and the magnetic-field dependent quantum interference effects (QIE) driving the conductance modulations, as well as the true contribution of the pseudospin degree of freedom, recently questioned by Winkler and Z\"ulicke~\cite{UZ}. Additionally, whereas weak localization in 2D is precursor of a true insulating state in the zero temperature limit, WAL points towards a robust metallic state, a fact in marked contrast with the usual scaling theory of localization for two-dimensional disordered systems~\cite{RMPLee}. 

Here, by means of an efficient quantum-transport simulation method, the weak-field magnetoconductance of disordered graphene-based materials is explored for the case of long-range disorder potential, 
which mimicks the effect of charges trapped in the underneath oxide layers. This study unveils the true contribution of pseudospin effects
in realistic disordered graphene. In two-dimensional graphene, a transition  from positive to negative magnetoconductance is confirmed to be related to the decay of the intervalley scattering contribution. 
Our study provides however a clarification of the microscopic origin of the reported experimental phase diagram~\cite{WALEX2} in terms of the dominant disorder characteristics, whereas weak antilocalization and pseudospin effects are found to be suppressed in the presence of graphene edges (graphene nanoribbons).

\begin{figure}
\onefigure[scale=0.66]{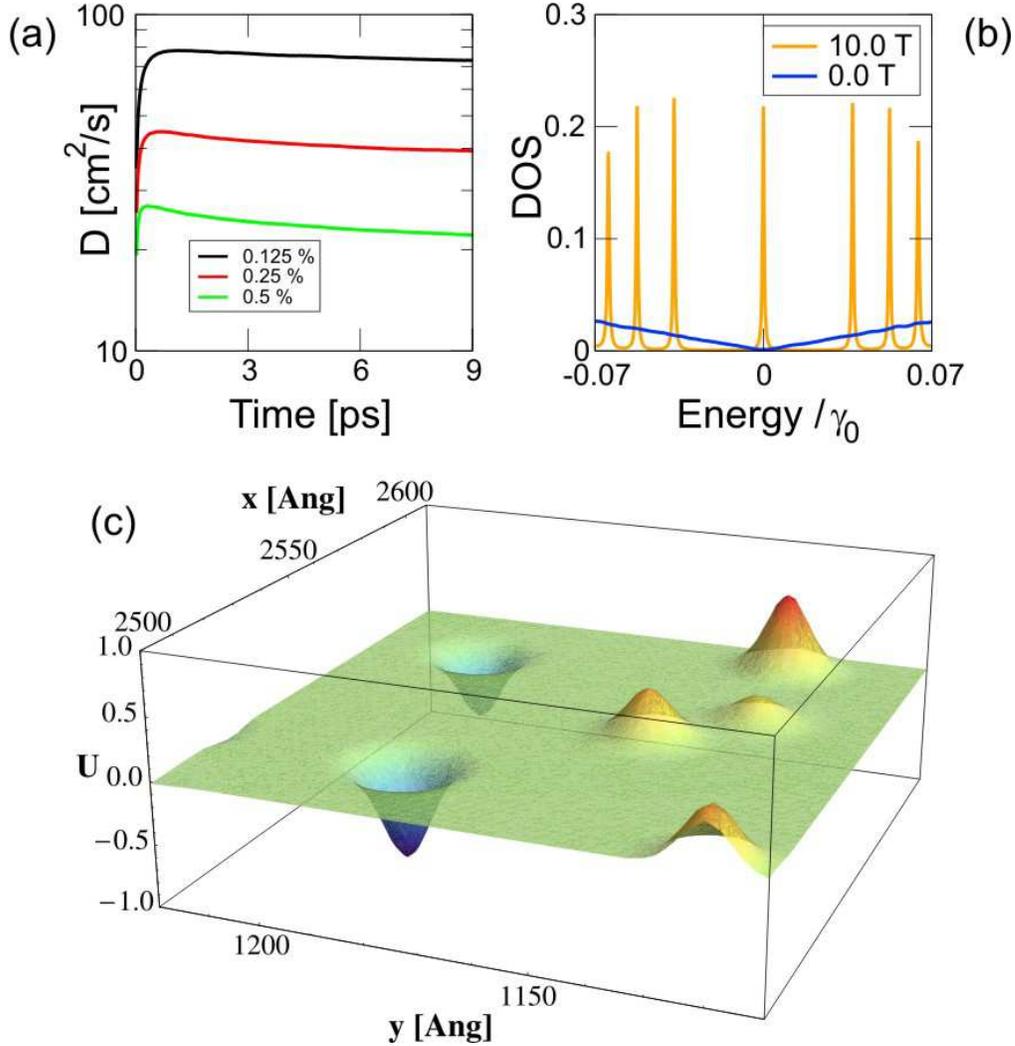}
\caption{(a) Time-dependent diffusion coefficient at Dirac point for several impurity densities and $W=2$.(b) Density of states of pristine graphene at zero and at 10 Tesla. (c) Schematic of the disordered potential in graphene due to long-range Coulomb impurities.}
\label{fig1}
\end{figure}

We use the $\pi$ electrons model of graphene well described by the tight-binding Hamiltonian ${\cal H} =  \sum_{\alpha}V_{\alpha}|\alpha \rangle\langle \alpha
|+\- \gamma_0 \sum_{\langle\alpha,\beta\rangle } e^{-i\varphi_{\alpha
\beta}} \ket{\alpha}\bra{\beta}$, with $\gamma_{0}=-2.7$ eV the hopping
integral between the nearest neighbor carbon atoms with distance $a=0.142$
nm, whereas $V_{\alpha}$ are the onsite energies of $\pi$-orbitals and include the (long range) disorder potential profile. The effect of the magnetic field is introduced through the Peierls phase~\cite{Peierls}, with a flux per hexagon being equal to  $\phi=\oint{\bf A}\cdot d{\bf l}=h/e\sum_{\rm hexagon}\varphi_{\alpha\beta}$. We implement a convenient gauge where $\sum_{\rm hexagon}\varphi_{\alpha\beta}$ can take integer multiples of $1/(N_{x}N_{y})$ to vary the magnetic field strength
and where $N_{x}$ and $N_{y}=N_{x}+1$ give the sample size. The total density of states in clean graphene at zero field and at $B=10$ Tesla is found to exhibit the well-known $\sqrt{n B}$ Landau level structure in perfect agreement  with theory~\cite{MCC_PRL104}.
We use a common model to describe the long-range disorder~\cite{LRDBB,GaussianLRD2}, which is given by the contributions from $N_{I}$ impurities randomly distributed at ${\bf
r}_{i}$ (among $N$ sites of the system)
with $V_{\alpha}=\sum_{i=1}^{N_{I}}\varepsilon_{i}\exp(-|{\bf
r}_{\alpha}-{\bf r}_{i}|^{2}/(2\xi^{2}))$, where $\xi$
defines the effective range, and $\varepsilon_{i}$ are chosen at random within $[-W/2, W/2]$ ($\gamma_{0}$-unit), such that $W$ gives the strength of the local potential profile (for an illustration see \ref{fig1} (c)). Different random configurations of graphene samples with same size, $\xi$, $W$, and $n_{i}= N_{i}/N$ constitute an
ensemble with given disorder strength. Following prior studies~\cite{LRDBB,GaussianLRD2}, we fix $\xi=3a=0.426$
nm as a typical value for a long-range potential, but vary $W$ to describe different screening situations.

The (magneto)-transport properties of large and disordered graphene
systems are simulated by using an efficient order-N Kubo method ~\cite{RocheM1}. From the study of the quantum wave-packet dynamics, the elastic mean free path $\ell_{e}(E)$ can be first derived from the time dependence of the diffusion  coefficient $D(E,t)$

$$D(E,t)=\frac{1}{t} \frac{ \sum_n \langle \Psi_n \lvert 
\delta(E-\hat{H}) \big{(}
\hat{U}^{\dag}(t) \hat{x} \hat{U}(t) - \hat{x} \big{)}2 \lvert \Psi_n
\rangle} { \sum_m \langle \Psi_m \lvert
 \delta(E-\hat{H}) \lvert \Psi_m \rangle }$$

with $\hat{x}$ the position operator along the transport direction ($x$), in the Schr\"{o}dinger representation. The behavior of $D(t)$ is related to the time evolution of $\lvert \Psi_n\rangle$ described by the operator $\hat{U}(t)=\Pi_{n=1}^{N_t}\exp(i\hat{H}\Delta t/\hbar)$ with $\Delta t$ the chosen time step. An efficient Chebyshev polynomial expansion method is used for $\hat{U}(t)$~\cite{RocheM1}. The calculations are performed for several initial wavepackets and for total elapsed computational times
$t=9 {\text{ps}}$, with $\Delta t = 2.44 {\text{fs}}$. The system size is about $(0.5\times0.6)\mu \text{m}^{2}$ (with about $1.15\times10^7$ carbon atoms) and periodic boundary conditions are applied. The Kubo conductivity is finally computed as $\sigma(E, t=N_t\Delta t)=e^{2}\rho(E)D(E,t)/4$, with $\rho(E)$ the total density of states and the magnetoconductance is calculated for
the total elapsed time.

In the case of disordered graphene nanoribbons (GNRs), we adopt the
Landauer-B\"{u}ttiker approach (within the Green functions formalism) to evaluate the transmission coefficient $T(E)$ and the related conductance. By performing a scaling analysis of these quantities~\cite{NLWhite} (averaged over a large number of different disorder configurations), we can extract the elastic mean free path $\ell_e$.

\begin{figure}
\onefigure[scale=0.60]{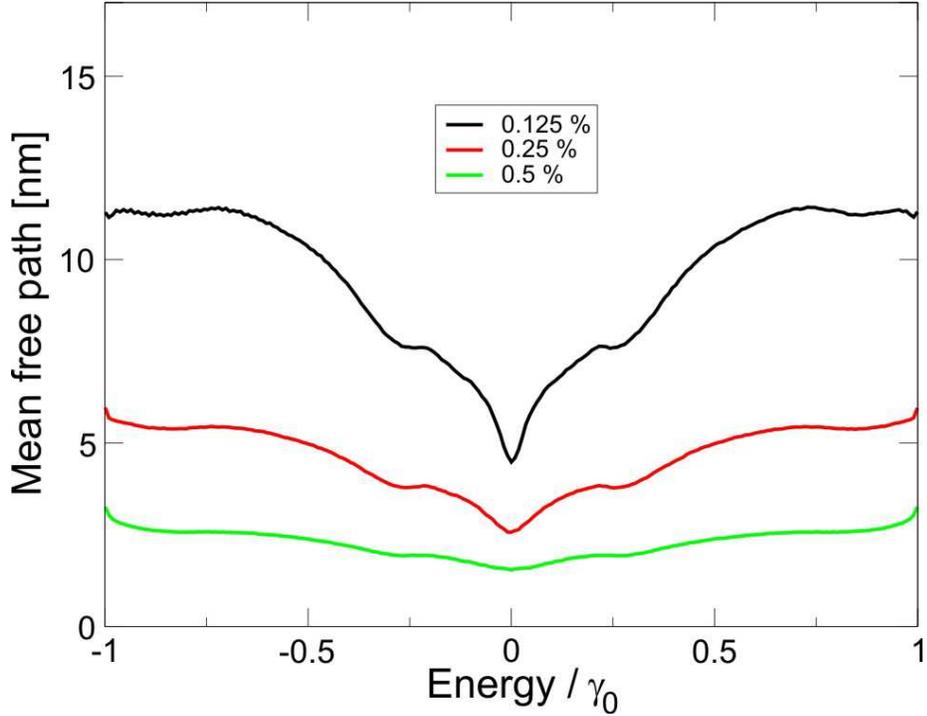}
\caption{Elastic mean free path at Dirac point for several impurity densities and $W=2$.}
\label{fig2}
\end{figure}

\ref{fig1} (a) shows the evolution of the wavepacket dynamics
(through time-dependent diffusion coefficients) at the Dirac point (E=0) for $W=2$ and increasing impurity density ($n_{i}=0.125\%, 0.25\%, 0.5\%$).  The diffusion coefficients are found to reach a saturation regime after about 1ps, and then to remain almost constant up to very long times, indicating a diffusive regime. At long elapsed propagation times, the saturation of the diffusion coefficient is followed by a time-dependent decay (for all energies and considered impurity densities), which pinpoints an increasing contribution of QIE and weak localization phenomena.

The maximum value of $D(t)=D_{\text{max}}$ allows for the evaluation of the elastic mean free path $\ell_{e}=D_{\text{max}}/2v_{F}$ (cf. \ref{fig2}) where $v_{F}=8.7\times 10^{5}{\text{ms}^{-1}}$ is the Fermi velocity. It is
seen that $\ell_{e}\sim 1/n_{i}$ in agreement with Fermi's Golden rule. The dependence of the elastic scattering time $\tau_{e}=\ell_{e}/v_{F}$ on the charge density $n$ follows $\tau_{e}\sim 1/\sqrt{n}$ for sufficiently small $n$ and $W$ (not
shown here).

\begin{figure}
\onefigure[scale=0.50]{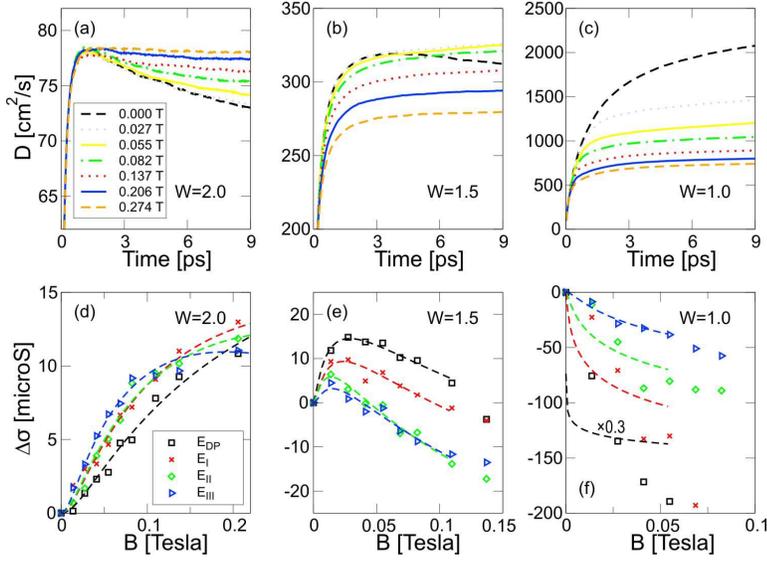}
\caption{Top panels: diffusion coefficient at the Dirac
point for various magnetic fields and $W$, $n_{i}=0.125\%$, $\xi=0.426$ nm. Bottom panels: $\Delta \sigma(B)$ for four different Fermi level positions ($E_{\text{DP}}=0, E_{\text{I}}=0.049$ eV, $E_{\text{II}}=0.097$ eV and $E_{\text{III}}=0.146$ eV). Dashed lines are fits as explained in the
text. For $W=1.0$ in (f) data and fit at $E_{\text{DP}}$ have been rescaled by 0.3 for clarity.}
\label{fig3}
\end{figure}

The effect of an applied magnetic field is shown in \ref{fig3}. Top
and bottom panels give the time-dependent diffusion coefficients at varying magnetic field and the magnetoconductance $\Delta \sigma(B)= \sigma(B)-\sigma(B=0)$, respectively. For $W=2$ (\ref{fig3}(a)), a steady suppression of QIE-related conductance
corrections is clearly seen with increasing $B$ up to $B=0.274$ T at which the decay of $D(t)$ has been fully suppressed. This is in accordance with the weak-localization phenomenon, as confirmed by the positive magnetoconductance reported in panel (d) for four selected energies (averages over 32 different configurations are performed). In contrast, the case $W=1.5$ shows a clearly different behavior as seen in \ref{fig3}(b) and (e) respectively for $D(t)$ and $\Delta \sigma(B)$ (for which 64 configurations have been averaged). Indeed, a change in the sign of $\Delta \sigma(B)$ is observed, in agreement with a crossover from weak localization to weak antilocalization, which depends not only on the field strength, but also the considered energy of charge carriers. We note that an estimation of the elastic mean free path gives $\ell_{e}\sim 9-20$nm, which is much smaller than our sample size.

For the case $W=1$, we first observe that at zero magnetic field, $D(t)$ does not reach a saturation regime  within our total computational time (dashed black curve in \ref{fig3}(c)). With increasing the magnetic field, $D(t)$ is found to be reduced and no onset of weak localization manifests. The corresponding $\Delta \sigma(B)$ curves displayed in panel (f) all exhibit a negative magnetoconductance, whose connection to a WAL regime, as discussed here below, is however not straightforward.

To rationalize all those results, we first note that the numerical study
reported in~\cite{GaussianLRD2} suggests that as long as $W\geq 1$, the long-range potential (with $\xi=0.426nm$) introduces strong valley mixing. This is consistent with our computed positive magnetoconductance for the case $W=2$ (\ref{fig3}(d)). By further decreasing the disorder strength (from $W=2$ to $W=1.5$), WAL occurs as seen in \ref{fig3}(e) which corresponds to the decay of intervalley scattering contribution (see also Figure 4 of ~\cite{GaussianLRD2}). Our study thus exhibits a transition from
weak localization (for $W=2$) to WAL ($W=1.5$), supporting that the enhancement of intervalley scattering contribution yields predominance of the weak localization effect. The analysis can be deepened by fitting the simulations using the phenomenological law
$\Delta \sigma(B)=e^{2}/\pi h\left\{{\cal
F}(\tau_{B}^{-1}/\tau_{\varphi}^{-1})-3{\cal F}(\tau_{B}^{-1}/
(\tau_{\varphi}^{-1}+\tau_{*}^{-1}))\right\}$, where ${\cal F}(z)=\ln
z+\psi(1/2+z^{-1})$, $\psi(x)$ is the digamma function
and $\tau_{B}^{-1}=4eDB/\hbar$~\cite{WAL2}. Here, for simplicity we restrict the model to a single additional elastic scattering time
$\tau_{*}$ (phenomenological parameter) that contains both contributions of intravalley and intervalley scattering~\cite{WAL2}.
The least-squares fits (dashed lines) are superimposed to the simulated $\Delta\sigma (B)$ (symbols), taking $\tau_{\varphi}=9{\rm ps}$
(our maximum computed time). For $W=2$ and $W=1.5$, $\tau_{*}$ ranges within $[1.1-2.3]{\rm ps}$ and $[1.5-6.3]{\rm ps}$, respectively (increasing values with increasing energy), thus confirming the weak localization regime for lowest $B$, which is fully consistent with \cite{WAL1} ($\tau_i<\tau_{\varphi}$). The case W=1 is more complicated, with a fit (with $\tau_{*}\geq\tau_{\varphi}$) only possible at lower magnetic fields and higher energies ($E_{II}$ and $E_{III}$) and out of reach for energies nearby the Dirac point.
Despite the obtained negative magnetoconductance, the interpretation in terms of weak antilocalization is actually not obvious.
Indeed, as seen on the diffusion coefficient behavior at zero magnetic field, the saturation regime is not reached within our maximum computed time, although the conduction clearly departs from a ballistic regime. By using $\ell_{e}=D_{\text{max}}/2v_{F}$, with an extrapolated maximum diffusion coefficient of $\geq 2500{cm}^2s^{-1}$, one gets $\ell_{e}\geq 330$nm, which is in the order of our graphene longitudinal size. Accordingly, the transport
regime is closer to a quasiballistic case, which jeopardizes the interpretation in terms of WAL.

\begin{figure}
\onefigure{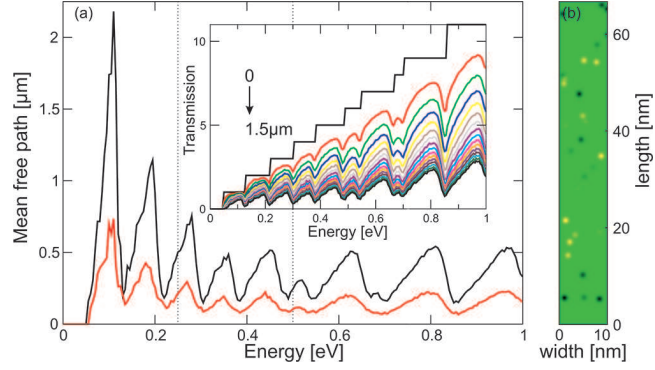}
\caption{(a) Inset: Transmission versus energy for a 10
nm-wide armchair ribbon (84-aGNR) with $W = 1/2$,  $n_{i} = 0.1\%$ and for different lengths (up to 1.5$\mu$m) of the disordered region. Main frame: Energy-dependent mean free path for $n_{i} = 0.1\%$ (black curve) and $n_{i} = 0.2\%$ (red curve). (b) Schematic of a short segment of a disordered nanoribbon with long-range scatters. Bright (dark) spots correspond to positive (negative) values of $V$.}
\label{fig4}
\end{figure}

To analyze the contribution of sample geometries (edges), we finally
simulate numerical transport in graphene nanoribbons~\cite{Ribbonpure1,NLWhite}. As a typical case, we focus on a 10 nm-wide armchair ribbon (84-aGNR) with impurity densities $n_{i} = 0.1\%, 0.2\%$, $\xi = 0.426$ nm and $W = 1/2$. The results for $T(E)$ are averaged over 100 disorder configurations (\ref{fig4}-inset). At the low fields (up to 0.5 T) considered here, the bandstructure does not change significantly with respect to the zero field case. Accordingly, the impact of the field on transport is only determined by the modifications of QIE.  \ref{fig4} (main frame) shows the elastic mean free path of disordered ribbons as a function of electron energy. As in the case of Anderson-type disorder or edge
defects~\cite{NLWhite}, $\ell_{e}$ exhibits important modulations
depending on the electron energy, with systematic decay around the
energies corresponding to the onset of new subbands. At low energies, $\ell_{e}$ can even reach a few microns for the considered disorder values.

\begin{figure}
\onefigure{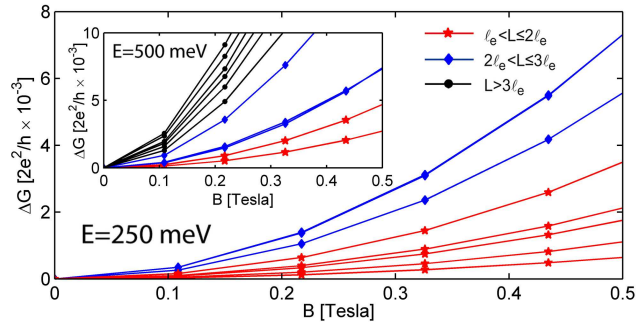}
\caption{ Magnetoconductance for the disordered ribbon
($W=0.5$) at two different selected energies (see \ref{fig4} vertical dashed lines) and for different ribbon lengths $L$. Different set of curves are identified depending on the ratio $L/\ell_{e}$.}
\label{fig5}
\end{figure}

\ref{fig5} shows the evolution of the quantity $\Delta G(B)=
G(B)-G(B=0)$ (for $n_{i} = 0.1\%$) for the same set of ribbon lengths
considered in \ref{fig4}, and at two different selected energies
(dashed vertical lines of \ref{fig4}). Note that the small value of $W=0.5$ considered here would prohibit valley mixing in the two-dimensional case. The ratio $L/\ell_{e}$ gives a suitable criterion to classify the transport regime as quasi-ballistic (when $L/\ell_{e}\leq1$) and diffusive or localized (when $L/\ell_{e}>1$). When $L/\ell_{e}>1$, a markedly positive magnetoconductance dominates ($\Delta G(B)>0$) in agreement with the standard weak localization regime (\ref{fig5}). The magnetoconductance variation increases for larger lengths, giving a stronger field-driven suppression of QIE. The results in the case $n_{i} = 0.2\%$ are analogous. Our numerical results evidence the absence of weak antilocalization for disordered GNRs because of the strong reintroduction of intervalley
scattering in presence of ribbon edge geometries.  One notes that the application of much larger magnetic fields has been recently theoretically predicted and experimentally found to yield very large magnetoresistance signals.~\cite{VLMR1,VLMR2}

In conclusion, in two-dimensional disordered graphene the crossover between weak localization and weak antilocalization was theoretically demonstrated to be driven by the nature of long-range Coulomb scattering potential. In contrast, predominance of weak localization effects was obtained in graphene nanoribbons. These results allow to bridge the nature of disorder in graphene with the localization phase diagram explored experimentally~\cite{WALEX2} and also open new directions for understanding pseudospin effects in graphene-based materials.

\acknowledgments
This work is supported by the NANOSIM-GRAPHENE project
(ANR-09-NANO-016-01) of ANR/P3N2009 and a Marie Curie Intra European
Fellowship within the 7th European Community Framework Programme.

\end{document}